
\input phyzzx
%
%
\newcount        \ObjClass
\chardef\ClassNum       = 0
\chardef\ClassMisc      = 1
\chardef\ClassEqn       = 2
\chardef\ClassRef       = 3
\chardef\ClassFig       = 4
\chardef\ClassTbl       = 5
\chardef\ClassThm       = 6
\chardef\ClassStyle     = 7
\chardef\ClassDef       = 8
\edef\NumObj    {\ObjClass = \ClassNum   \relax}
\edef\MiscObj   {\ObjClass = \ClassMisc  \relax}
\edef\EqnObj    {\ObjClass = \ClassEqn   \relax}
\edef\RefObj    {\ObjClass = \ClassRef   \relax}
\edef\FigObj    {\ObjClass = \ClassFig   \relax}
\edef\TblObj    {\ObjClass = \ClassTbl   \relax}

\edef\StyleObj  {\ObjClass = \ClassStyle \relax}
\edef\DefObj    {\ObjClass = \ClassDef   \relax}
%
%
\def\gobble      #1{}%
\def\trimspace   #1 \end{#1}%
\def\ifundefined #1{\expandafter \ifx \csname#1\endcsname \relax}%
\def\trimprefix  #1_#2\end{\expandafter \string \csname #2\endcsname}%
\def\skipspace #1#2#3\end%
    {%
    \def \temp {#2}%
    \ifx \temp\space \skipspace #1#3\end
    \else \gdef #1{#2#3}\fi
    }%
\def\stylename#1{\expandafter\expandafter\expandafter
    \gobble\expandafter\string\the#1}
\ifundefined {protect} \let\protect=\relax \fi
\def\checkchapterlabel%
    {%
    \catcode`\@=11
    {\protect\expandafter\if\chapterlabel\noexpand\relax
        \global\let\chapterlabel=\relax \fi}
    \catcode`\@=12
    }%
\begingroup
\catcode`\<=1 \catcode`\{=12
\catcode`\>=2 \catcode`\}=12
\xdef\LBrace<{>%
\xdef\RBrace<}>%
\endgroup
%
%
\newcount\equanumber \equanumber=0
\newcount\eqnumber   \eqnumber=0
\def\(#1)%
        {%
        \ifnum \equanumber<0 \eqnumber=-\equanumber
            \advance\eqnumber by -1 \else
            \eqnumber=\equanumber\fi
        \ifmmode\ifinner(\eqnum {#1})\ifdraft{\rm[#1]}\fi%
        \else \expandafter\ifx\csname Eq_#1\endcsname \relax
        \eqno(\eqnum {#1})\ifdraft{\rm[#1]}\fi \else(\eqnum {#1})\fi\fi
        \else(\eqnum {#1})\fi\ifnum%
            \equanumber<0 \global\equanumber=-\eqnumber\global\advance%
            \equanumber by -1\else\global\equanumber=\eqnumber\fi}%
\def\eqnum #1%
    {%
    \LookUp Eq_#1 \using\eqnumber\neweqnum
    {\rm\label}%
    }%
\def\neweqnum #1#2%
    {%
    \checkchapterlabel%
    {\protect%
        \xdef\eqnoprefix{\ifundefined{chapterlabel}\else\chapterlabel.\fi}}%
    \ifmmode \xdef #1{\eqnoprefix #1}%
        \else\message{Undefined equation \string#1 in non-math mode.}%
             \string#1 \let #1=\relax%
             \global\advance \eqnumber by -1 %
        \fi%
    \EqnObj \SaveObject{#1}{#2}%
    }%
\everydisplay = {\expandafter \let\csname Eq_\endcsname=\relax
                 \expandafter \let\csname Eq_?\endcsname=\relax}%
%
%
\newcount\tablecount \tablecount=0
\def\Table  #1{Table~\tblnum {#1}}%
\def\tblnum #1{\TblObj \LookUp Tbl_#1 \using\tablecount
        \SaveObject \label\ifdraft [#1]\fi}%
\def\tbldef #1{\TblObj \SaveContents {Tbl_#1}}%
\def\inserttable #1#2#3%
    {%
    \tbldef {#1}{#3}\goodbreak%
    \midinsert
        \smallskip
        \hbox{\singlespace
                   \hskip 0.4in
                   \vtop{\parshape=2 0pt 361pt 58pt 303pt
                        \noindent{\bf\Table{#1}}.\enspace #3}
                   \hfil}
        #2
        \smallskip%
    \endinsert
    }%
%
%
\newcount\figurecount \figurecount=0
\def\Figure #1{Figure~\fignum {#1}}%
\def\fig    #1{fig.~\fignum {#1}}%
\def\fignum #1{\FigObj \LookUp Fig_#1 \using\figurecount
     \SaveObject \label\ifdraft [#1]\fi}%
\def\figdef #1{\FigObj \SaveContents {Fig_#1}}%
\def\figlist  {\FigObj \ListObjects}%
\def\insertfigure #1#2#3%
    {%
    \figdef {#1}{#3}%
    \midinsert
        \smallskip
        #2
        \hbox{  \singlespace
                \hskip 0.4in
                \vtop{\parshape=2 0pt 361pt 65pt 296pt
                      \noindent{\bf\Figure{#1}}.\enspace #3}
                \hfil}
        \smallskip%
    \endinsert
    }%
%
%
\newcount\theoremcount \theoremcount=0
\def\fbf#1{\expandafter\ifx\csname Thm_#1\endcsname\relax \bf\fi}
%
%
%
%
%
%
%
%
%
%
%
%
%
\newcount\referencecount \referencecount=0
\newcount\refsequence   \refsequence=0
\newcount\lastrefno     \lastrefno=-1
\def\refsymbol#1{[\refrange#1-\end]}%
\def\[#1]%
        {\refsymbol{#1}}
\def\^[#1]#2{
        \if.#2\rlap.\attach{\refsymbol{#1}}\let\refendtok=\relax%
        \else\if,#2\rlap,\attach{\refsymbol{#1}}\let\refendtok=\relax%
        \else\attach{\refsymbol{#1}}\let\refendtok=#2\fi\fi%
        \discretionary{}{}{}\refendtok}%
\def\refrange #1-#2\end%
    {%
    \refnums #1,\end
    \def \temp {#2}%
    \ifx \temp\empty \else -\expandafter\refrange \temp\end \fi
    }%
\def\refnums #1,#2\end%
    {%
    \def \temp {#1}%
    \ifx \temp\empty \else \skipspace \temp#1\end\fi
    \ifx \temp\empty
        \ifcase \refsequence
            \or\or ,\number\lastrefno
            \else  -\number\lastrefno
        \fi
        \global\lastrefno = -1
        \global\refsequence = 0
    \else
        \RefObj \edef\temp {Ref_\temp\space}%
        \expandafter \LookUp \temp \using\referencecount\SaveObject
        \global\advance \lastrefno by 1
        \edef \temp {\number\lastrefno}%
        \ifx \label\temp
            \global\advance\refsequence by 1
        \else
            \global\advance\lastrefno by -1
            \ifcase \refsequence
                \or ,%
                \or ,\number\lastrefno,%
            \else   -\number\lastrefno,%
            \fi
            \label
            \global\refsequence = 1
            \ifx\suffix\empty
                \global\lastrefno = \label
            \else
                \global\lastrefno = -1
            \fi
        \fi
        \refnums #2,\end
    \fi
    }%
%
%
%
%
\def\refdef #1{\RefObj \SaveContents {Ref_#1}}%
\def\reflist  {\RefObj \ListObjects}%
%
%
%
%
\newif\ifSaveFile
\newif\ifnotskip
\newwrite\SaveFile
\let\IfSelect=\iftrue
\edef\savefilename {\jobname.aux}%
\def\Def#1#2%
    {%
    \expandafter\gdef\noexpand#1{#2}%
    \DefObj \SaveObject {#2}{\expandafter\gobble\string#1}%
}%
\def\savestate%
    {%
    \ifundefined {chapternumber} \else
        \NumObj \SaveObject {\number\chapternumber}{chapternumber} \fi
        \ifundefined {appendixnumber} \else
        \NumObj \SaveObject {\number\appendixnumber}{appendixnumber} \fi
    \ifundefined {sectionnumber} \else
        \NumObj \SaveObject {\number\sectionnumber}{sectionnumber} \fi
    \ifundefined {pagenumber} \else
        \advance\pagenumber by 1
        \NumObj \SaveObject {\number\pagenumber}{pagenumber}%
        \advance\pagenumber by -1 \fi
    \NumObj \SaveObject {\number\equanumber}{equanumber}%
    \NumObj \SaveObject {\number\tablecount}{tablecount}%
    \NumObj \SaveObject {\number\figurecount}{figurecount}%
    \NumObj \SaveObject {\number\theoremcount}{theoremcount}%
    \NumObj \SaveObject {\number\referencecount}{referencecount}%
    \checkchapterlabel
    \ifundefined {chapterlabel} \else
        {\protect\xdef\chaplabel{\chapterlabel}}
        \MiscObj \SaveObject \chaplabel {chapterlabel} \fi
    \ifundefined {chapterstyle} \else
        \StyleObj \SaveObject {\stylename{\chapterstyle}}{chapterstyle} \fi
    \ifundefined {appendixstyle} \else
        \StyleObj \SaveObject {\stylename{\appendixstyle}}{appendixstyle}\fi
}%
\def\Contents #1{\ObjClass=-#1 \SaveContents}%
\def\Define #1#2#3%
    {%
    \ifnum #1=\ClassNum
        \global \csname#2\endcsname = #3 %
    \else \ifnum #1=\ClassStyle
        \global \csname#2\endcsname\expandafter=
        \expandafter{\csname#3\endcsname} %
    \else \ifnum #1=\ClassDef
        \expandafter\gdef\csname#2\endcsname{#3} %
    \else
        \expandafter\xdef \csname#2\endcsname {#3} \fi\fi\fi %
    \ObjClass=#1 \SaveObject {#3}{#2}%
    }%
\def\SaveObject #1#2%
    {%
    \ifSaveFile \else \OpenSaveFile \fi
    \immediate\write\SaveFile
        {%
        \noexpand\IfSelect\noexpand\Define
        {\the\ObjClass}{#2}{#1}\noexpand\fi
        }%
    }%
\def\SaveContents #1%
    {%
    \ifSaveFile \else \OpenSaveFile \fi
    \BreakLine
    \SaveLine {#1}%
    }%
\begingroup
    \catcode`\^^M=\active %
\gdef\BreakLine %
    {%
    \begingroup %
    \catcode`\^^M=\active %
    \newlinechar=`\^^M %
    }%
\gdef\SaveLine #1#2%
    {%
    \toks255={#2}%
    \immediate\write\SaveFile %
        {%
        \noexpand\IfSelect\noexpand\Contents
        {-\the\ObjClass}{#1}\LBrace\the\toks255\RBrace\noexpand\fi%
        }%
    \endgroup %
    }%
\endgroup
\def\ListObjects #1%
    {%
    \ifSaveFile \CloseSaveFile \fi
    \let \IfSelect=\GetContents \ReadFileList #1,\savefilename\end
    \let \IfSelect=\IfDoObject  \input \savefilename
    \let \IfSelect=\iftrue
    }%
\def\ReadFileList #1,#2\end%
    {%
    \def \temp {#1}%
    \ifx \temp\empty \else \skipspace \temp#1\end \fi
    \ifx \temp\empty \else \input #1 \fi
    \def \temp {#2}%
    \ifx \temp\empty \else \ReadFileList #2,\end \fi
    }%
\def\GetContents #1#2#3%
    {%
    \notskipfalse
    \ifnum \ObjClass=-#2
        \expandafter\ifx \csname #3\endcsname \relax \else \notskiptrue \fi
    \fi
    \ifnotskip \expandafter \DefContents \csname #3_\endcsname
    }%
\def\DefContents #1#2{\toks255={#2} \xdef #1{\the\toks255}}%
\def\IfDoObject #1#2%
    {%
    \notskipfalse \ifnum \ObjClass=#2 \notskiptrue\fi \ifnotskip \DoObject
    }%
\def\DoObject #1#2%
    {%
    \ifnum \ObjClass = \ClassTbl        \par\noindent Table~#2.
    \else \ifnum \ObjClass = \ClassFig  \par\noindent Figure~#2.
    \else \item {#2.}
    \fi\fi
    \ifdraft\edef\temp {\trimprefix #1\end}[\expandafter\gobble \temp]~\fi
    \expandafter\ifx \csname #1_\endcsname \relax
        \ifdraft\relax\else\edef\temp {\trimprefix #1\end}%
        [\expandafter\gobble \temp]\fi%
    \else
        \csname #1_\endcsname
    \fi
    }%
\def\OpenSaveFile   {\immediate\openout\SaveFile=\savefilename
                     \global\SaveFiletrue}%
\def\CloseSaveFile  {\immediate\closeout\SaveFile \global\SaveFilefalse}%
\OpenSaveFile\CloseSaveFile 
%
%
\def\LookUp #1 #2\using#3#4%
    {%
    \expandafter \ifx\csname#1\endcsname \relax
        \global\advance #3 by 1
        \expandafter \xdef \csname#1\endcsname {\number #3}%
        \let \newlabelfcn=#4%
        \ifx \newlabelfcn\relax \else
            \expandafter \newlabelfcn \csname#1\endcsname {#1}%
        \fi
    \fi
    \xdef \label  {\csname#1\endcsname}%
    \gdef \suffix {#2}%
    \ifx \suffix\empty \else
        \xdef \suffix {\expandafter\trimspace \suffix\end}%
        \xdef \label  {\label\suffix}%
    \fi
    }%
%
%
%
\newcount\appendixnumber        \appendixnumber=0
\newtoks\appendixstyle          \appendixstyle={\Alphabetic}
\newif\ifappendixlabel          \appendixlabelfalse
\ifundefined{numberedchapters}  \fi
\def\APPEND#1{\par\penalty-300\vskip\chapterskip\spacecheck\chapterminspace
        \global\chapternumber=\number\appendixnumber
        \global\advance\appendixnumber by 1
        \chapterstyle\expandafter=\expandafter{\the\appendixstyle}
\chapterreset
        \titlestyle{Appendix\ifappendixlabel~\chapterlabel\fi.~ {#1}}
        \nobreak\vskip\headskip\penalty 30000}
\def\append#1{\APPEND{#1}}

%
%
%
\def\references#1{\par\penalty-300\vskip\chapterskip\spacecheck
        \chapterminspace\line{\fourteenrm\hfil References\hfil}
        \nobreak\vskip\headskip\penalty 30000\reflist{#1}}
\def\figures#1{\par\penalty-300\vskip\chapterskip\spacecheck
        \chapterminspace\line{\fourteenrm\hfil Figure Captions\hfil}
        \nobreak\vskip\headskip\penalty 30000\figlist{#1}}

%
%
\newif\ifdraft\draftfalse
\newcount\yearltd\yearltd=\year\advance\yearltd by -1900
\def\draft{\ifdraft\relax\else\drafttrue
        \def\draftdate{preliminary draft:
                \number\month/\number\day/\number\yearltd\ \ \hourmin}%
        \paperheadline={\hfil\draftdate} \headline=\paperheadline
        {\count255=\time\divide\count255 by 60 \xdef\hourmin{\number\count255}
                \multiply\count255 by-60\advance\count255 by\time
                \xdef\hourmin{\hourmin:\ifnum\count255<10 0\fi\the\count255} }
        \message{draft mode}\fi }
%

%

\overfullrule=0pt
\Pubnum={IASSNS--HEP--91/57}
\date{October 1991}
\titlepage
\title{ Charged Black Holes In Two-Dimensional String Theory }
\author{Michael D. McGuigan,\foot{Research supported in part by the Department
        of Energy, contract DE-FG02-90ER-40542. Present address: University of
        Florida, Dept.\ of Physics, Gainesville FL 32611 USA.}
Chiara R. Nappi\foot{Research supported in part by
        the Ambrose Monell Foundation.}}
\address{Institute for Advanced Study, School of Natural Sciences\break
         Princeton, New Jersey 08540 USA}
\andauthor{Scott A. Yost\foot{Research supported in part by the Department
        of Energy, contract DE-FG05-86ER-40274. Present address: University of
        Tennessee, Dept.\ of Physics, Knoxville TN 37996 USA.}}
\address{University of Florida, Department of Physics\break
         Gainesville, Florida 32611 USA}

\abstract{ We discuss  two dimensional string theories containing gauge
fields, introduced
either via  coupling to open strings, in which case we get a Born-Infeld type
action,  or  via heterotic compactification.  The solutions
of  the modified background field equations
 are charged black holes which exhibit
 interesting
space time geometries. We also compute their masses and charges. }
\endpage

\chapter{Introduction}
Recently, black hole solutions to two-dimensional string theory have attracted
much interest\[Witten,Mandal,blackholes,ILS, Dealw, Kumar]. The black hole
discussed so far is a
solution of the bosonic closed string theory
on the sphere, and it is neutral.
It is interesting to ask what happens to the black hole
in string theories which include a gauge field, permitting it to have a
charge.  It is also interesting to ask how loop corrections to the string
theory modify the conformally invariant tree level solutions. Both of these
modifications appear naturally in the open string case, since the gauge field
couples to gravity through loop effects\[ACNY,CLNY1]. Therefore
 we  compute the corrections to the black hole solution
which come from  adding holes and crosscaps to the
closed string worldsheet and coupling a gauge field to the boundary.
As in \[ACNY,CLNY1], we work in the approximation of a
slowly-varying electromagnetic field, and the effective action
for the gauge field is then of the
Born-Infeld type. We solve the beta functions of  the two
dimensional string theory that are valid at the leading order in $\alpha'$,
  and find a charged black hole solution.
 For completeness and comparison we also discuss
a two dimensional heterotic string solution, which also gives rise to a
 charged black hole. Of course in this case the gauge field
comes in at the tree level. However the black hole solution is of
the same general type as those obtained from loop corrections.

This paper is organized as follows. In section 2 we review the black
hole solution of two dimensional string theory, and show how it is
modified in the presence of generic loop corrections. In section 3
we derive the specific modification to the string tree level
equation of motion that arise from coupling gauge fields to open strings.
As in the critical string,
these modifications can be derived by standard background field calcutions,
 but they can really be understood only
if one interpret them as loop corrections. Therefore, in section 4,
as in ref.~\[CLNY2],  we actually perform a loop calculation
by building the boundary operator for holes and crosscaps, and requiring the
BRST anomaly of the tree level state to cancel against that of the
 boundary operator.
 This calculation is similar to that of ref.~\[CLNY2], but somehow
 more subtle since we are in two dimensions.
The results are the same as those from background field calculations.
 In section 5, we find an
 exact solution for the Born-Infeld black hole.
In section 6 we discuss and solve the heterotic theory. In section 7,
 we discuss the geometric interpretation of these
black holes. In some cases, the solutions look very
similar to Reissner-Nordstrom black holes, although the parameters in the
solution often have a different interpretation. For example, the role
of the charge in the Reissner-Nordstrom solution may be played by the
cosmological constant instead.
For some ranges of parameters, the charged
black hole for the open string
bears no resemblance to the Reissner-Nordstrom form, and
one finds that the Penrose diagram tiles the plane with black holes.
Finally, in section 8, we derive the  masses and charges in the various cases.

\chapter{The black hole solution}
In ref.~\[Witten] it was proved that the black hole  is a solution
of an exact conformal field theory,
namely the WZW model with gauge group $SL(2,R)/U(1)$. However a black
 hole solution
can be derived also by solving the two-dimensional beta function equations  for
   the
graviton-dilaton system at the leading order in $\alpha'$, as in
ref.~\[Mandal].
  To the leading order in $\alpha'$,
the beta functions of the closed bosonic string  in $D$ dimensions are\[CFMP]
$$\eqalignno{
\beta^G_{\mu\nu} = R_{\mu\nu} &+ 2\nabla_\mu\nabla_\nu\Phi = 0 &\(uno) \cr
\beta^\Phi =  -R - 4\nabla^2 \Phi & + 4(\nabla\Phi)^2 + c = 0  &\(unoprimo)
\cr}$$
where $ c = 2(D-26)/3\alpha'$. We are assuming here that the tachyon field
is zero, and drop it from the equations. Our curvature conventions
follow ref.~\[MTW], as in ref.~\[CFMP].

We look for a metric of the form
$$
G_{tt} = -g(r),\qquad G_{rr} = {1\over g(r)} \(Gansatz)
$$
where $t$ is a timelike coordinate and $r$ is spacelike,
and assume that $\Phi(r)$ is static.
Then the graviton beta functions become
$$\eqalignno{
\beta^G_{tt} &= {g\over2}(g'' -2 g'\Phi') & \(betagt)\cr
\beta^G_{rr} &= 2 \Phi'' - {1\over 2g}(g'' - 2 g'\Phi')
& \(betagr)\cr} $$
and the dilaton beta function becomes
$$
\beta^\Phi = g'' + 4g(\Phi')^2 - 4g\Phi'' - 4 g'\Phi' + c,
\(betaphi)$$
where primes denote derivatives with respect to $r$.
Also, $\beta^G_{tr}=0$ identically.

Together, \(betagt) and \(betagr) imply that $\Phi'' = 0$, so the dilaton
background takes the form
$$ \Phi = \Phi_0 - {Qr\over 2}. \(five)$$
Then a linear combination of \(betagt) and \(betaphi) gives
$$ \beta^\Phi-{2\over g}\beta^G_{tt} = Qg' + Q^2g + c = 0, \(six)$$
with the family of solutions
 $g(r)= -(c Q^{-2} + 2m e^{-Qr})$. Requiring the metric to be asymptotically
flat for $r\rightarrow\infty$ fixes $Q = \sqrt{-c}$.
It was shown in ref.~\[Witten] that this solution is the two dimensional
version of the four dimensional Schwarzschild solution.
 (Notice that this solution would be trivial if $c=0$.)
The free parameter $m$ is proportional to the mass, as described in sect.~8.

String loops can modify the beta functions\[Love,FS]. To zeroth order in
derivatives of the fields,
the effect is simply to add a cosmological constant term
$$
        S_{\rm loop} = \int d^{D} x \sqrt{-G} \Lambda e^{2(n-1)\Phi} \(cosmo)
$$
to the effective action, weighted
by the dilaton factor appropriate for $n$ handles\[FS,CLNY1].
{}From these corrections to the effective
action, we can infer corrections to the beta functions.
Specifically, \(uno) and \(unoprimo) take the form
$$\eqalign{
R_{\mu\nu} + 2\nabla_\mu\nabla_\nu\Phi &= {n\over2}\Lambda G_{\mu\nu}
e^{2n\Phi}
\cr -R -4(\nabla^2\Phi) + 4(\nabla\Phi)^2 + c &= (1-n)\Lambda e^{2n\Phi},\cr}
\(seven)$$
where $n$ is chosen to represent the leading nonvanishing loop correction.

Then the r.h.s.~of \(six) becomes $\Lambda e^{2n\Phi_0} e^{-nQr}$, and
the generic solution is
$$
g(r) = \cases{1 - 2m e^{-Qr} + b_n e^{-nQr} & if $n\ne 1$\cr
              1 - 2m e^{-Qr} + b_1 Qr e^{-Qr} & if $n = 1$\cr}
\(loopg)$$
where
$$
b_n = {\Lambda\over c\;(n-1)}\;e^{2n\Phi_0} {\rm\ \ for\ }n\ne 1,\qquad
b_1 = - {\Lambda\over c}\;e^{2\Phi_0}. \(bns)
$$
The most general loop-generated dilaton potential is a linear combination of
the terms on the r.h.s.~of eq.~\(seven), and the black hole metric includes
a sum over the $b_n$ terms in eqs.~\(loopg) -- \(bns).
 It is interesting to point out that \(loopg) implies that in the presence
of loop corrections flat space is no longer allowed as a solution.

  We will show that open strings  obey nonlinear Born-Infeld electrodynamics.
For hole and crosscap-type corrections\[CLNY2,P+C] in the absence
of a gauge field,  it will turn out that $n=1/2$ in the above formulas.
Open strings with a weak electric charge but no
cosmological constant will  have $n=3/2$.

Heterotic strings have charged solutions with an electromagnetic field
which will mimic loop corrections with $n=2$,
 so they also fit into this general class of
corrections to the black hole.
The first closed string loop corrections (torus-level) will have $n=1$.
Although this case is very interesting, since closed string loop corrections
appear in every string theory, we will not discuss it here any further, except
in section 8 when we compute masses.

\chapter{Open strings in two dimensions}

The original means of introducing a gauge field into string theory was to
couple it to charges on the ends of an open string\[ChanPaton].
In this section, we will
derive the open string beta functions for the $D=2$ string, and use them to
find the equations which must be satisfied by a black hole coupled to an
abelian gauge field.

We first consider the simplest case, that of a $U(1)$ electromagnetic field.
The combined open and closed string worldsheet action is\[CLNY1]
$$\eqalign{
I =&\ {1\over 4\pi\alpha'}\int d^2 z
\sqrt{-\gamma}\left\{ \gamma^{ab}G_{\mu\nu}(X)
                \partial_a X^\mu \partial_b X^\nu
 + \alpha' R^{(2)}\Phi(X) + {\alpha'\over\epsilon^2}T(X)\right\}\cr
&+ {1\over 2\pi\alpha'}\oint ds\left\{ A_\mu(X){\partial\over\partial s}X^\mu
 + \alpha' k \Phi(X) + {\alpha'\over\epsilon}\Theta(X) \right\}\cr} \(wsaction)
$$
where $T$ and $\Theta$ are respectively the closed and open string
``tachyons\rlap.'' The bare tachyon couplings include factors of a worldsheet
cutoff $\epsilon$ with dimensions of length\[tachyonbeta].
The gauge field $A_\mu$ has been rescaled by a factor of
$2\pi\alpha'$, so that $F_{\mu\nu}$ is dimensionless.
The term containing the extrinsic worldsheet curvature
$k$ is needed to complete the coupling of the dilaton to the Euler density.

Demanding conformal invariance of the worldsheet action \(wsaction) requires
two new beta functions,
$\beta^A_\mu$ and $\beta^\Theta$, to vanish. This implies new equations
of motion, which to lowest order in derivatives of the gauge fields are given
by
$$\eqalignno{
&\beta^A_\mu = (G - F^2)^{-1}_{\lambda\nu}\nabla^\nu F_{\mu}{}^{\lambda}
        + \nabla^\nu\Phi\; F_{\nu\mu} = 0 & \(betaA)\cr
\beta^\Theta =&\ (G-F^2)^{-1}_{\mu\nu}\nabla^\mu\nabla^\nu \Theta -
\nabla^\mu\Phi\nabla_\mu\Theta + {1\over \alpha'}\Theta = 0. &   \(betaTheta)
\cr}$$
Up to $\Theta$-dependent terms in eq.~\(betaA),
these are the variational equations of an open string contribution
$$
S_{\rm open} = -\kappa\int d^2 x  e^{-\Phi}\sqrt{-{\rm det}(G+F)}
        \Bigl\{1+ \alpha'(G-F^2)^{-1}_{\mu\nu}
        \nabla^\mu \Theta \nabla^\nu \Theta - \Theta^2
        \Bigr\} \(openaction)
$$
to the effective action,
where $\kappa$ is a positive open-string coupling constant
with dimensions of inverse length-squared.\foot{If $\kappa$ were negative,
the gauge field would couple to gravity with the wrong sign. This is
a consequence of our curvature conventions, which follow refs.\ \[MTW,CFMP].}
The gauge-field
part of eq.~\(openaction) is the Born-Infeld action for non-linear
electrodynamics\[BI,ACNY]. The presence of boundary fields in
\(wsaction) does not affect
 \[ACNY, CLNY1] the beta functions \(uno) and \(unoprimo),
so that the closed string tree-level effective action\[CFMP,CG] remains
$$
S_{\rm closed} = \int d^2 x\sqrt{-G}e^{-2\Phi}\left\{ R + 4(\nabla\Phi)^2 -
 (\nabla T)^2 + {4\over\alpha'} T^2 -c\right\}.
\(closedaction)$$

Like the tachyon
contributions to the graviton and dilaton beta functions\[tachyonbeta],
the $\Theta$ terms in $\beta^A$ require a sum over all loop orders in the
sigma model.
We will not pursue this here, but simply note that
eqs.~\(betaA) -- \(openaction) can be
expected to receive higher-order corrections in $\Theta$, just as the
closed string action contains nontrivial higher-order tachyon interactions.
The tachyons in the effective action are renormalized with respect to the bare
tachyon backgrounds in the worldsheet action\[tachyonbeta].

As in the closed string case, we will consider a linear dilaton
background of the form \(five). Then
the field redefinitions $T = e^{\Phi}{\widehat T}$,
$\Theta = e^{\Phi/2}{\widehat\Theta}$ remove the negative mass-terms from
the tachyon equations of motion, showing that both ``tachyons''
are actually massless
when $D=2$, in agreement with the expected spectrum\[spec].
The massless open and closed string tachyons are the physical
degrees of freedom of the
$D=2$ open string. The metric, dilaton, and gauge field are non-dynamical
backgrounds. We will look for black hole solutions without tachyon backgrounds,
and drop $T$ and $\Theta$ in the following.

The total space-time effective action associated
 with \(wsaction) is obtained \[CLNY1] by adding
 \(openaction) to \(closedaction). Then the equation of motion
derived from total action are

$$\eqalign{
&R_{\mu\nu} + 2\nabla_\mu\nabla_\nu\Phi + {\kappa\over 4}e^{\Phi}
        \sqrt{\det(1 + G^{-1}F)}\left({G+F^2\over G-F^2}\right)_{\mu\nu} = 0\cr
&-R + 4(\nabla\Phi)^2 - 4\nabla^2\Phi + c + {\kappa\over2}e^{\Phi}
        \sqrt{\det(1+ G^{-1}F)} = 0\cr} \(loopybetas)
$$
The above results are based on background field calculations.
  They rely  on an expansion
in derivatives of the fields, whose momenta are therefore supposed
to be small. However, due to the linear dilaton background, and the
fact that in two dimensions, the backgrounds are massive in the
worldsheet sense
(except the tachyon), such an expansion might be unjustified.
In the next section we will show  that the corrections to the beta functions
inferred from putting together the effective actions \(openaction) and
 \(closedaction) can be interpreted as loop-corrections due to the
 the effect of small holes and crosscaps in the worldsheet.
This loop calculation
is of interest of its own merits, since it involves massive modes of the
string,
and is a nice check of the results of
background field calculations.

\chapter{Loop corrections from the boundary operator}
   The basic ideas and methods of our derivation in this section
 are the same as in the critical case
\[CLNY1,Love,FS,CLNY2,P+C],
but with  modifications due mostly to the linear dilaton background.
Namely, we impose  the BRST invariance of the closed string field state
and get the string tree level equations of motion \(uno) and \(unoprimo),
as in \[SZ]. Then we represent the insertion of a hole in the worldsheet
via a boundary operator \[CLNY2,P+C], and require the BRST invariance
of the sum of both the closed string tree-level state and the boundary
operator.
 This will give the loop corrected equations as in \[CLNY1, CLNY2, CLNY3]

In this calculation, we will use a weak field expansion for the closed string
backgrounds.
We first derive the tree level beta function starting with the string state
$$\ket{\Psi}= \Bigl\{{\widehat T}(x) + {\widehat h}_{\mu\nu}(x)
\alpha^{\mu}_{-1}\;{\widetilde{\alpha}}^{\;\nu}_{-1} +
{\widehat\phi}(x)(c_{-1}{\widetilde b}_{-1}
- b_{-1}{\widetilde c}_{-1})\Bigr\}c_1{\widetilde c}_1\ket{iQ/2} \(vertex)
$$
where $\ket{iQ/2} = e^{-Q\cdot x/2}\ket{0}$ is the proper Liouville
gravitational dressing factor for $D=2$ strings\[DK]. The fields
${\widehat h}_{\mu\nu}$ and ${\widehat \phi}$ are perturbations around
the background graviton and dilaton,
and the hats are a reminder that they are
 related to those in the sigma model lagrangian by a field
redefinition, yet to be determined.
The \hbox{$c_1{\widetilde c}_1 e^{-Q\cdot x/2}$} factor in \(vertex)
is what converts the $SL_2(R)$ vacuum $\ket{0}$ to a puncture vacuum,
appropriate for attaching a fixed vertex\[CLNY2,P+C,DK].

Before discussing the BRST invariance conditions, it is useful to write
the mode expansion of the coordinates $X^\mu$.
The closed string coordinates may be expanded as\[GSW]
$$
X^\mu(\sigma,\tau) = x^\mu + 2\alpha'P^\mu \tau + i\sqrt{\alpha'\over2}
        \sum_{n\ne0}{1\over n}\left\{
        \alpha^\mu_n e^{-2in(\tau-\sigma)} + {\widetilde\alpha}_n^\mu
        e^{-2in(\tau+\sigma)}\right\} . \(expansion)$$
The linear dilaton background makes the definition of $P^\mu$
somehow subtle.
The operator $X^\mu$ is hermitian, so $P^\mu$ must be hermitian as well.
If $\ket{0}$ is the ordinary $SL_2(R)$ vacuum and \hbox{$\ket{p} \equiv
e^{ip\cdot X(0)}\ket{0}$}, then the linear dilaton background
implies\[FMS,GSW] that there is a background
momentum: \hbox{$\VEV{p+iQ|p} = 1$}. Consequently the momentum in
eq.~\(expansion) must be shifted by half the background charge, so that
\hbox{$P^\mu\ket{p} = (p-iQ/2)^\mu\ket{p}$}.

The Virasoro generators are now
$$\eqalignno{
&L_n^X = {1\over2}\sum_m \eta_{\mu\nu} :\alpha_m^\mu \alpha_{n-m}^\nu:
- {in\over2}\sqrt{\alpha'\over2}Q_\mu\alpha^\mu_n - {D\over 24}\delta_{n0}
&\(Xvir)\cr
&L_n^{\rm ghost} = \sum_m (m-n):c_m b_{n-m}: + {1\over12}\delta_{n0}
&\(ghvir)\cr}$$
and the BRST charge is
$$
Q_{\rm BRST} = \sum_n c_{-n}\left(L^X_n + \coeff12 L_n^{\rm ghost}\right).
\(QBRST)
$$
Here, $\alpha_0^\mu = \sqrt{\alpha'/2}P^\mu$.
The additional $Q$-dependent terms are from the contribution of the
linear dilaton background to the stress tensor. Such terms were originally
found for Liouville theory in ref.~\[ChodosThorn].
These terms are analogous
to ones appearing for bosonized ghosts, for the same reason\[ThornSFT,GSW].
The ghost part is unchanged from the critical case, and the right-movers
are defined analogously.
Applying the techniques of ref.~\[CLNY2] with the modified Virasoro operators,
we find that $(Q +{\widetilde Q})_{\rm BRST} \ket{\Psi}=0$ implies
$$\eqalign{
&{\alpha'\over 4}\nabla^2 {\widehat h_{\mu\nu}} - {\widehat h_{\mu\nu}} = 0,\cr
&{\alpha'\over 4}\nabla^2{\widehat \phi} -{\widehat \phi} = 0.\cr}
\(treelevel) $$
The BRST gauge conditions coming from $L_{\pm 1}$ terms in $Q_{\rm BRST}$ are
$$\eqalign{
\left(\partial+\coeff12 Q\right)^{\nu} {\widehat h}_{(\mu\nu)} &-
\left(\partial-\coeff12 Q\right)_{\mu}{\widehat\phi} = 0\cr
\left(\partial+\coeff12 Q\right)^{\nu} {\widehat h}_{[\mu\nu]} &= 0,\cr}
\(gauge)
$$
where parentheses and backets around the indices denote symmetrization
and antisymmetrization, respectively.
These equations need to be compared with the linearized version of
\(uno) and \(unoprimo). To this purpose, as in \[CLNY2],
we will use a weak-field expansion
$$\eqalign{
        &G_{\mu\nu}(X) = \eta_{\mu\nu} + h_{\mu\nu}(X)\cr
        &\Phi(X) = \Phi_0 - \coeff12 Q_\mu X^\mu + \phi(X)\cr}\(weakfield)
$$
with $Q_\mu = Q\delta_{\mu r}$ as in  \(five).
Substituting
\(weakfield) into the beta functions \(uno) and \(unoprimo), we find
$$\eqalignno{
\beta^G_{\mu\nu} = -\coeff12(\partial&+\coeff12 Q)^2 h_{\mu\nu} + \coeff18
   Q^2  h_{\mu\nu} + \coeff12 (\partial_\mu J_\nu + \partial_\nu J_\mu)
& \(betaGlinear)\cr
\beta^\Phi + \eta^{\mu\nu}\beta^G_{\mu\nu} &= (\partial+\coeff12 Q)^2\phi
        - \coeff14{Q^2}\phi -J_\mu(\partial^\mu\phi -\coeff12 Q^\mu)
& \(betaPhilinear)\cr }$$
where
$$
J_\mu = (\partial+Q)^\nu h_{\mu\nu} - \coeff12 \partial_\mu h^\nu{}_\nu
        + 2\partial_\mu\phi. \(bigJ)
$$

Fixing the gauge $J_\mu = 0$ and making the field redefinitions
$$\eqalign{
&{\widehat h}_{\mu\nu} = e^{\Phi_0-\Phi}h_{\mu\nu},\cr
&{\widehat\phi} = e^{\Phi_0-\Phi}(\coeff12 h^\nu{}_\nu - 2\phi).\cr} \(redef)
$$
 reproduces eq.\ \(treelevel), while the gauge condition
$J_\mu=0$ turns into \(gauge).
Note that ${\widehat\phi}$ is a linear combination of
the dilaton vertex and trace of the graviton\[SZ, CLNY2],
and the $e^{\Phi_0-\Phi}$ scaling is the same as for the tachyon. This scaling
 is associated with the Euler characteristic of a puncture.

Although we have not discussed the antisymmetric tensor before, we may note
that the gauge condition \(gauge) on the antisymmetric part of
${\widehat h}_{\mu\nu}$  implies that the antisymmetric part of
 eq.~\(treelevel) is identically zero (remember ${Q^2\over
4}={4\over\alpha'}$),
 so that at the tree level
 the  antisymmetric tensor beta function vanishes identically.
This has important consequences, which we will come back to.

We now compute the loop corrections. The boundary state
that describes the insertion of a hole on the worldsheet with abelian gauge
field attached is \[CLNY1,CLNY2,CLNY3]
$$\eqalign{
\ket{B}_F = \lambda e^{\Phi_0}&\sqrt{-{\rm det}[\eta(x)+F(x)]}\cr
&\times \exp\left\{-\sum_{n=1}^\infty {1\over n}\Bigl({\eta-F\over \eta+F}
        \Bigr)_{\mu\nu}\alpha_{-n}^\mu{\widetilde\alpha}_{-n}^\nu\right\}
   (c_0+{\tilde c}_0) \ket{iQ/2}_X\ket{B}_{\rm ghost} \cr}   \(Bstate)
$$
since $\ket{iQ/2}$ satisfies the zero-mode condition $P^\mu = 0$.
The presence of a background momentum makes this reminiscent of the boundary
states for bosonized ghosts\[CLNY2] and superghosts\[yost].
The ghost boundary state is the same as in the critical case\[CLNY2].
The open string coupling constant $\lambda$ will be proportional to $\kappa$
in \(openaction).
The state $\ket{B}$ is attached to a worldsheet via a cylinder. We need to
integrate over the modulus (length) $\tau$ of the attached cylinder,
which  has the effect of
multiplying $\ket{B}$ by a closed string propagator,
and also need to
insert the required ghost zero-modes (which eventually will
combine with those in \(Bstate) to reproduce the appropriate vacuum \[CLNY2]).
  Therefore the complete
state for inserting a boundary in the worldsheet is\[CLNY2]
$$\ket{F}=
-(b_0 + {\widetilde b}_0)\int_0^\infty d\tau e^{-\tau(L_0 + {\widetilde L}_0)}
\ket{B}_F =
-(b_0 + {\widetilde b}_0)(L_0 + {\widetilde L}_0)^{-1}\ket{B}_F. \(totalB)
$$
 We need to remember \[CLNY2] that  the
state $\ket{B}$ itself is BRST-invariant by virtue of its boundary conditions,
and that
$$\{(Q + {\widetilde Q})_{\rm BRST}, b_0 + {\widetilde b}_0 \} =
 L_0 + {\widetilde L}_0. \(comm)$$
Therefore commuting $(Q+{\widetilde Q})_{\rm BRST}$
past the ghost insertion in \(totalB),
produces an inverse propagator, which cancels the propagator in \(totalB).
This cancellation presents no problem for the massive modes,
 and for them the correction
to the tree level equations \(treelevel) can be read directly from the boundary
operator, and no divergence calculation is involved. (This is not too
striking. In the BRST approach we derived directly \(treelevel)
 without computing
any divergence, while in the sigma model approach the same equations
emerged from a beta function calculation). However the cancellation is more
problematic for the massless modes, since in this case one gets a zero over
zero ambiguity which can be dealt with by some regularization
procedure, for instance giving the boundary state an off-shell
momentum as in \[CLNY3].

 Therefore, requiring as in critical case\[CLNY2] that
the {\it sum} of \(vertex) and \(totalB)
be BRST-invariant one  gets here for the massless tachyon
$$
{\alpha'\over4}\nabla^2 {\widehat T}  =
\lambda e^{\Phi_0}\sqrt{-\det(\eta+F)} \(Tcorrect)
$$
including the $F$-dependence.
This is completely analogous to  the derivation of the
loop corrections for massless modes in the critical string.

Similarly one derives for the massive modes the loop corrected  equations
$$\eqalign{
{\alpha'\over4} \nabla^2 {\widehat h}_{\mu\nu} - {\widehat h}_{\mu\nu} &=
        -\lambda e^{\Phi_0}\sqrt{-\det(\eta+F)}
        \Bigl({\eta+F^2\over\eta-F^2}\Bigr)_{\mu\nu}\cr
{\alpha'\over4} \nabla^2 {\widehat\phi}-{\widehat\phi} &=
        \lambda e^{\Phi_0}\sqrt{-\det(\eta+F)}.
   \cr}
        \(GDcorrect)
$$
while the gauge condition \(gauge) receives no loop corrections.

The loop corrections \(GDcorrect) are exactly what
are needed to explain the extra terms in the graviton and dilaton beta
functions which can be inferred from the open string effective
action \(openaction) ( here we restrict ourselves to the case $\Theta = 0$).
 Indeed, \(GDcorrect)
agree with the linearized equations of motion \(loopybetas)
 if $\kappa = 8\lambda/\alpha'$. This agreement confirms the validity
of our procedure to obtain loop correction to massive, as well as massless
states.
In \(Tcorrect) ${\widehat T}$ is the massless tachyon, which is related to the
sigma model tachyon by ${\widehat T} = {T}e^{\Phi_0-\Phi}$. This relation
 is consistent with the tachyon field
redefinition discussed earlier. (Subtracting $\Phi_0$ in the field
redefinition is optional. It is done to maintain the connection
between $e^{\Phi}$ and the string loop coupling.)
The tachyon correction \(Tcorrect) may be accounted for by adding a term
$$
-\kappa e^{\Phi}T(X)\sqrt{-\det(G+F)}     \(linearinT)
$$
to the effective lagrangian, whose variational equation cancels the r.h.s.
of \(Tcorrect).

A remark is in order here.
On the r.h.s.\ of \(GDcorrect), we have symmetrized the matrix
$\left({\eta-F\over\eta+F}\right)_{\mu\nu}$ which comes from the boundary
state. Its antisymmetric part may be considered to be the correction to the
equation for the antisymmetric tensor field $B_{\mu\nu}$.
However, as noted earlier, since its field strength $H_{\lambda\mu\nu}$
vanishes identically in $D=2$, the closed string
tree-level equations are independent of $B_{\mu\nu}$. Therefore
 the antisymmetric part of the boundary corrections must vanish too,
which can happen only if $F=0$.

The same conclusion can be reached also in the approach of section 3.
Indeed in the presence
of a boundary, the antisymmetric tensor enters the beta functions and effective
action via the substitution \hbox{$F_{\mu\nu} \rightarrow F_{\mu\nu} +
B_{\mu\nu}$} everywhere\[CLNY1]. Varying the action with respect to
$B_{\mu\nu}$
gives the antisymmetric tensor beta function, which turns out to imply
\hbox{$F + B = 0$}. Therefore, the gauge field can be gauged away.

This problem can be solved by considering a nonabelian
gauge group instead, and choosing a background in an abelian subgroup.
Orientable open strings can couple to $U(N)$, while
unorientable ones can couple to $SO(N)$ or $Sp(N)$, where in the latter case,
$N$ is even\[opengroup]. In the unorientable case, the antisymmetric tensor
is simply absent. In the orientable case, it will eat a $U(1)$ factor by
the process described above, leaving behind an $SU(N)$ group and no
antisymmetric tensor.

In the case of nonorientable strings,
one needs to take in account also  the cross-cap contribution which appears
 at the same order in the loop expansion
as the boundary state. The insertion of a cross-cap on the world-sheet
can be represented by the operator \[CLNY2,P+C]
$$
\ket{C} =  2^{D/2}\eta\lambda e^{\Phi_0}
        \exp\left\{\sum_{n=1}^{\infty}(-)^{n+1}{1\over n}
                \eta_{\mu\nu}\alpha^\mu_{-n}{\widetilde\alpha}^{\;\nu}_{-n}
                \right\} \ket{iQ/2}_X \ket{C}_{\rm ghost}  \(xcap)
$$
which will also contribute to \(GDcorrect).

 If one takes the background
field in an abelian subgroup, the final form of the effective action
is indeed (omitting the tachyon)
$$
S_{\rm open} = -\kappa\int d^D x\; e^{-\Phi}\left\{\Tr\sqrt{-\det(G + F)}
        + \eta 2^{D/2}\sqrt{-G} \right\} \(nonabelian)
$$
where $D=2$ for our purposes, and $\eta = 0$, $-1$, or $1$ when the group
is $SU(N)$, $SO(N)$ or $Sp(N)$, respectively.
The trace is over the group indices, which are suppressed, and the
term proportional to $2^{D/2}$ comes from including from the crosscap
contribution.

It is useful to choose a particular abelian subgroup, so that we can
evaluate the trace in \(nonabelian).
We will take the background to be in an abelian subgroup
with a single generator $\lambda^a$. The group elements
are represented by $N\times N$ matrices, and we normalize the generators
by \hbox{$\Tr \lambda^a\lambda^b = 2\delta^{ab}$}. The generator $\lambda^a$
of the abelian subgroup
can be chosen so that
$$
\Tr 1 = N,\qquad \Tr F^{2n} = 2F^{2n},\qquad \Tr F^{2n+1} = 0. \(traces)
$$
Using \(traces)
and the fact that $\det(G+F)$ is even in $F$ (since $F$ is antisymmetric)
we rewrite the open string part of the action as
$$
S_{\rm open} = -\kappa\int d^2 x\; e^{-\Phi}\left\{2\sqrt{-\det(G+F)} +
         (N-2+2\eta)\sqrt{-G}   \right\}.    \(NAaction)
$$
The equations of motion derived by adding \(NAaction) to\(closedaction)
are now
$$\eqalign{
&R_{\mu\nu} + 2\nabla_\mu\nabla_\nu\Phi + {\kappa\over 2}e^{\Phi}
        \sqrt{\det(1 + G^{-1}F)}\left({G+F^2\over G-F^2}
        \right)_{{\hskip-3pt}\mu\nu} {\hskip-6pt}
        + {\kappa\over 4} (N+2\eta-2) e^{\Phi}G_{\mu\nu} = 0\cr
&-R + 4(\nabla\Phi)^2 - 4\nabla^2\Phi + c + {\kappa}e^{\Phi}
        \sqrt{\det(1+ G^{-1}F)} + {\kappa\over2}(N+2\eta-2)e^{\Phi} = 0.\cr}
\(finalbetas)
$$
Comparing eqs.\ \(NAaction) and \(finalbetas) for $F=0$ with the general
forms \(cosmo) and \(seven) shows that the open string contributes a
cosmological constant $\Lambda = -\kappa(N+2\eta)$,
which vanishes when the
gauge group is $SO(2)$.
We will solve \(finalbetas)  in sect.~5.

\chapter{Open string Born-Infeld black holes}

We will now search for a charged black hole solution to the background
field equations \(finalbetas) for the open $SU(N)$,
$SO(N)$, or $Sp(N)$ string, with an abelian
background $F_{\mu\nu}$.
In two dimensions, the field strength has just one independent component
\hbox{$F_{tr} = - F_{rt} = f$}.
The gauge field beta function \(betaA) is
$$
\beta^A_r = (1-f^2)^{-1}f' - \Phi' f.   \(betaA1)
$$
Note that $\beta^A_t$ vanishes identically.

\def\khat{\widehat\kappa}
Rewriting the beta functions \(finalbetas) using the metric
\(Gansatz) gives the equations
$$\eqalign{
-\coeff12 g'' + 2g'\Phi' = -\coeff12 g'' + 2g'\Phi' +2g\Phi'' &=
-{\kappa\over2}e^{\Phi}\left\{{1+f^2\over\sqrt{1-f^2}}
        +{N\over2}+\eta-1\right\}\cr
g'' + 4g(\Phi')^2-4g\Phi'' - 4g'\Phi' + c
&=\ -{\kappa}e^{\Phi}\left\{\sqrt{1-f^2}+{N\over2}+\eta-1 \right\}
\cr}\(sy56)
$$
for the metric and dilaton background.
The graviton beta functions again imply that that $\Phi'' = 0$,
so the ansatz \(five) is still appropriate. The equations to solve are then
$$\eqalignno{
(f^2)' &= -Qf^2(1-f^2) &\(sy7)\cr
g' + Q g &= -{c\over Q} - {2\khat\over Q}e^{\Phi}
        \left\{(1-f^2)^{-1/2} +{N\over2}+\eta-1\right\}.
&\(tracebetag)\cr}
$$

The solution to \(sy7) must be positive and vanish at infinity. A family of
solutions is
$$
f^2(r) = q^2(e^{Qr}+q^2)^{-1}. \(fsoln)
$$
Substituting this into \(tracebetag) gives
$$
\left(g(r)e^{Qr}\right)' = -{c\over Q}e^{Qr} + {\khat\over Q}
        (2-2\eta-N)e^{Qr/2}
        - {\khat\over Q}\sqrt{e^{Qr}+q^2}
\(sy9)$$
where $\khat = \kappa e^{\Phi_0}$.
Integrating \(sy9) and using $Q^2 = -c$ gives
$$\eqalign{
g(r) =&\ 1 - 2m e^{-Qr} + {2\khat\over c}(N+2\eta-2) e^{-Qr/2}\cr
&+ {4\khat q\over c}e^{-Qr}
\left\{\sqrt{q^{-2}e^{Qr}+1}\ -\ \sinh^{-1}(qe^{-Qr/2})\right\}.\cr}
\(sy10)$$

The curvature is then
$$\eqalign{
R = &-2mce^{-Qr} + \left({N\over2}+\eta-1\right)\khat e^{-Qr/2}\cr
 &\ + {\khat}(e^{Qr}+q^2)^{-1/2} - 4\khat qe^{-Qr} \sinh^{-1}(qe^{-Qr/2}).\cr}
\(sy11)$$
This has a singularity at $r = -\infty$, which
is where the usual curvature singularity occurs.
Event horizons occur when $g(r) = 0$. In general there
will be more than one, as for the usual charged black hole.

To first order in $q^2$, the black hole \(openaction) corresponds to a solution
coupled to ordinary linear electrodynamics with a cosmological constant,
$$
g(r) = 1 - 2m e^{-Qr} + {2\khat\over c}(N+2\eta)e^{-Qr/2}
 - {2\khat q^2\over c} e^{-3Qr/2}. \(linsoln)
$$
When $q=0$, the pure cosmological constant correction is present, and has
the form \(loopg) with $n = 1/2$.
The cosmological constant $\Lambda = -\kappa(N+2\eta)$
may be cancelled by coupling nonorientable strings to $SO(2)$, as noted in the
previous section.

The $U(1)$ string has no gauge field, due to the fact that the antisymmetric
tensor $B_{\mu\nu}$ acts as a Lagrange multiplier. However, it still has
a cosmological constant, which the action \(openaction) shows to be
$\Lambda = -\kappa$. In that case, the modified black hole metric is of the
form \(loopg) for $n=1/2$, specifically,
$$
g(r) = 1 - 2me^{-Qr} + {\khat\over c}e^{-Qr/2}. \(purecc)
$$

\chapter{Black holes in two dimensional heterotic string theory}

Heterotic strings provide an alternative way to couple a gauge field
to string theory.
In this section, we find solutions of the form of a two dimensional
charged black hole to a string theory heterotically compactified to
two dimensions.  In the two non compact space-time dimensions
the left-moving sector will be taken to be bosonic,
and the right-moving sector supersymmetric, so that
the heterotic sigma model coupling the string
to the bosonic backgrounds of interest is\[CFMP]
$$\eqalign{
I = {1\over 2\pi\alpha'}\int d^2z d\theta \Bigl\{ &G_{\mu\nu}(X)DX^\mu
\widetilde{\partial} X^\nu -
 4\alpha'(D\widetilde {\partial}\sigma)\Phi(X)\cr
&+ A_{\mu a}(X)D X^\mu \widetilde{\partial}J^a\Bigr\} \cr } \(hetsigma)
$$
Where $J^a$  is the current of the internal gauge group, whose nature
will be discussed shortly. Above we have used superfield notations
for the fields in the right sector and the world-sheet conformal gauge
$\gamma_{ab}= e^{2\sigma}\delta_{ab}$.
We will assume that the background $A_\mu$ takes values in an abelian
subgroup of the full gauge group, as in the open string case.

The metric, dilaton, and gauge field beta functions
at string tree level and to the leading order in
$\alpha'$ are given by\[CFMP]
$$\eqalign{
\beta^G_{\mu\nu} &= R_{\mu\nu} + 2\nabla_{\mu} \nabla_{\nu} \Phi
-\coeff12 F_{\mu}{}^{\lambda} F_{\nu\lambda}\cr
\beta^{\Phi} &= \coeff14 F^2 -R + c
+ 4 (\nabla \Phi )^2 - 4\nabla^2 \Phi\cr
\beta^A_{\nu} &= \nabla_{\mu} F_{\nu}{}^{\mu} -2 \nabla_{\mu} \Phi
                F_{\nu}{}^{\mu}\cr} \(hetbeta)
$$
These beta functions can all be derived from the effective action
$$
S = \int d^2x \sqrt{-G} e^{-2 \Phi}\Bigl\{R+4(\nabla \Phi )^2
- c -\coeff14 F^2\Bigr\}.     \(Shet)
$$
Unlike the heterotic string in critical dimensions, the central charge $c$
will be nonzero here, which is crucial for the existence of two dimensional
black hole solutions, as in the bosonic case.
The actual value of $c$ may differ from the bosonic case,
 and will be given later. It
is the only way in which \(Shet) depends on the specific details of
the heterotic string construction.
The antisymmetric tensor background contributes only through its field
strength $H_{\lambda\mu\nu}$, which vanishes in two dimensions, so it has
been omitted.

We look for a metric of the form \(Gansatz) and a dilaton background \(five).
The requirement that the beta functions vanish then reduces to
$$\eqalignno{
-g''-Qg' + f^2 &= 0  &   \(mm1) \cr
\left(fe^{-2 \Phi}\right)' &= 0  &  \(mm2) \cr}$$
where we have written $F_{tr} = -F_{rt} = f$. The second equation
can be easily solved by $f(r) = \sqrt{2} Qqe^{-Qr}$.
Then the first equation becomes
$$
-g'' - Qg' + 2Q^2 q^2 e^{-2Qr} =0.     \(mm3) 
$$
This is a special case of the equations considered in sect.~2 with
solution
$$
g= 1-2me^{-Qr} +q^2e^{-2Qr}.       \(gsoln)
$$
As usual, asymptotic flatness requires $c = -Q^2$.
Notice that \(mm2) tells that a combination of the gauge field and the dilaton
field is a constant. This seems to be a general feature of these two
dimensional
models, and a similar relation holds also in the Born-Infeld case as shown in
\(fsoln).
The solution  \(gsoln) can have two horizons and is qualitatively
different from the
charged black hole of ref.~\[ILS], which was a solution of a $D=2+1$
string theory compactified to two dimensions. The global structure
of this solution as well as that of the half-loop open string solution is
discussed in the next section.

We need to describe the gauge group and internal space which gives rise to
a heterotic theory of the type described above.
Heterotic string compactifications with a linear dilaton background
and tree level cosmological constant
have been constructed in\[dAPS],
but in {\it four} non-compact dimensions. Non-critical superstrings have also
been discussed in \[KS, KRT]. Our construction differs in that we choose
a GSO projection which allows the tachyon to be present. For two
non-compact dimensions this mode turns out to be massless, as in the
bosonic string, and poses no problem.

Requiring that
the total central charge of the left (right) ghost, non-compact space-time,
dilaton background and internal field theories vanish yields the
conditions
$$\eqalign{
-26 +2 + {3\alpha'\over 2}Q^2 + \widetilde{c}_I &= 0\cr
-15 +3 + {3\alpha'\over 2}Q^2 + c_I &= 0, \cr }  \(deltac)
$$
where $(\tilde{c_I},c_I)$ are the (left,right) central charges of the
internal theory.
 Subtracting the two equations in \(deltac), we get $ \widetilde{c}_I
- c_I =12$. We take the internal charges to be
$\widetilde{c}_I = 12,  c_I = 0$.
Comparison with \(deltac) then shows that
$ c = - Q^2 = -{8\over\alpha'}.$
 Our construction   could be interpreted as a heterotic construction with a
 two dimensional
superstring on the right and a 14-dimensional bosonic string on the left
compactified down to two dimensions and giving rise to a rank 12 group.
 Here we take the internal theory on the left
to consist of an  $E_8 \times SO(8)$ or $SO(24)$ current algebra.

To be consistent at loop level, the appropriately GSO projected
 theory must be modular invariant.
The approach we follow is to start with
 a left-right symmetric superstring
theory with worldsheet fermions on the left and right, that is
modular invariant. Then, by using
Gepner's construction \[Gepner], we will then obtain a modular invariant
 heterotic theory   with the left worldsheet fermions
replaced by the fermionic representation of a group current algebra.

We imitate the construction of
refs.\ \[DH,SW], where new types of ten dimensional superstring theories
were found by imposing a new GSO projection which projects out
the spin ${3\over 2}$ gravitino. Our  theory will not be supersymmetric
and will contain a tachyon, but in $D=2$
that state will turn out to be massless. We begin as in refs.\ \[DH,SW],
 by choosing a modular invariant set of boundary conditions
 in $\sigma$ and $\tau$
which consists of summing over the same boundary conditions for
left and right movers. Namely, we project the (NS,NS) sector
onto states with $(-)^{F+\widetilde{F}}=1$ and add a (R,R)
sector also projected onto states with $(-)^{F+\widetilde{F}}=1$.
In $d = D-2$ transverse dimensions  one obtains three
sectors (NS,NS), $(\overline{\rm NS},\overline{\rm NS})$ and (R,R),
where $\overline{NS}$ refers to $NS$ states of odd fermion number.
The partition function in the left-right symmetric
theory with world-sheet fermions on the left and right is then
$$\eqalign{
\Tr &\left[  \coeff12 (1+(-)^F)e^{2\pi i\tau M^2}\right]_{NS}
\Tr \left[\coeff12 (1+(-)^{\widetilde{F}})
e^{2\pi i\tau \widetilde{M}^2}\right]^{*}_{NS}\cr
&\quad + \Tr\left[\coeff12(1-(-)^F)e^{2\pi i\tau M^2}\right]_{\overline{NS}}
\Tr \left[\coeff12 (1-(-)^{\widetilde{F}})
e^{2\pi i\tau \widetilde{M}^2}\right]^{*}_{\overline{NS}}\cr
&\quad + \Tr \left[\coeff12(1+(-)^F)e^{2\pi i\tau M^2}\right]_{R}
\Tr \left[\coeff12 (1+(-)^{\widetilde{F}})
e^{2\pi i\tau \widetilde{M}^2}\right]^{*}_{R}\cr} \(partitiontraces)
$$
where $M^2, \widetilde{M}^2 $ are the right and left $(mass)^2$ operators.

Performing the traces in $d=D-2$ transverse dimensions, we obtain
the modular invariant partition function
$$\eqalign{ 2^{d-2}
\tau_2^{-{d\over 2}}{1\over |\theta_1'|^d}
\Bigl\{(\theta_3^{d\over 2} + \theta_4^{d\over 2})
& (\theta_3^{d\over 2} + \theta_4^{d\over 2})^{*}
+
(\theta_3^{d\over 2} - \theta_4^{d\over 2})
(\theta_3^{d\over 2} - \theta_4^{d\over 2})^{*}
+
2(\theta_2^{d\over 2})
(\theta_2^{d\over 2})^{*}\Bigr\}\cr
&= 2^{d-1}
\tau_2^{-{d\over 2}} {1\over |\theta_1'|^d}
(|\theta_3|^{d} + |\theta_4|^{d}+|\theta_2|^{d}).\cr} \(partitionthetas)
$$

Now one can  obtain a heterotic construction from the
above theory following \[Gepner]. The point is that
the above expressions
\hbox{$\theta_3^{d\over 2} + \theta_4^{d\over 2}$},
\hbox{$\theta_3^{d\over 2} - \theta_4^{d\over 2}$} and
$\theta_2^{d\over 2}$ are proportional to the
characters of the $SO(d)$ helicity group current algebra in the scalar,
vector and spinor representation (denoted repectively by
$b_0, b_v$ and $b_s + b_{\overline{s}}$). It turns out that the characters
of an $E_8\times SO(8+d)$ or $SO(24+d)$ current algbra in the scalar,
vector and spinor representation (denoted respectively
by $B_0, B_v$ and $B_s + B_{\overline{s}}$) transform in exactly
the same way as the helicity characters provided one replaces
$b_0$ by $B_v$, $b_v$ by $B_0$ and $b_s + b_{\overline{s}}$
by $-B_s - B_{\overline{s}}$. This amounts to projecting onto states
with $(-)^{F+\widetilde{F}}=-1$ in the $NS$ sector
of the fermionic formulation of $E_8\times SO(8+d)$
or $SO(24+d)$. The partition function in the
$SO(24+d)$ heterotic string theory case is then given by
$$\eqalign{
\Tr &\left[ \coeff12 (1+(-)^F)e^{2\pi i\tau M^2}\right]_{NS}
\Tr\left[\coeff12 (1-(-)^{\widetilde{F}})
e^{2\pi i\tau \widetilde{M}^2}\right]^{*}_{NS,SO(24+d)}\cr
&\quad + \Tr\left[\coeff12(1-(-)^F)e^{2\pi i\tau M^2}\right]_{\overline{NS}}
\Tr\left[\coeff12 (1+(-)^{\widetilde{F}})
e^{2\pi i\tau \widetilde{M}^2}\right]^{*}_{\overline{NS},SO(24+d)}\cr
&\quad - \Tr\left[\coeff12(1+(-)^F)e^{2\pi i\tau M^2}\right]_{R}
\Tr\left[\coeff12 (1+(-)^{\widetilde{F}})
e^{2\pi i\tau \widetilde{M}^2}\right]^{*}_{R,SO(24+d)}\cr}\(partitionSO24)
$$
Note that the overall minus sign for the $(R,R)$ sector indicates that this
is the fermionic portion of the partition function. Performing the
traces in eq.~\(partitionSO24), the partition function becomes
$$\eqalign{ 2^{2 + {2d\over3}}
\tau_2^{{-d\over 2}} &
 {1\over |\theta_1'|^d}{1\over {{\theta_1'}^4}^{*}}
((\theta_3^{d\over 2} + \theta_4^{d\over 2})
(\theta_3^{12+{d\over 2}}- \theta_4^{12+ {d\over 2}})^{*} \cr
&+ (\theta_3^{d\over 2} - \theta_4^{d\over 2})
(\theta_3^{12 + {d\over 2}} + \theta_4^{12+{d\over 2}})^{*}
+ 2(\theta_2^{d\over 2}) (-\theta_2^{12+ {d\over 2}})^{*})\cr}\(partitionfinal)
$$
and is modular invariant. Again the GSO projection amounts taking an extra
minus sign for $({-1})^{\widetilde{F}}$ in the $NS$ sector of the fermionic
formulation of $SO(24+d)$.
For the special case of $D=2$, with no transverse dimensions, the partition
function is given by
$$
{8\over {\theta_1'}^{4*}}(\theta_3^{12} - \theta_4^{12} - \theta_2^{12})^{*}
. \(partitionD=2)
$$
The first two theta functions give the bosonic partition function,
and the last one is the fermionic partition function.

It can be shown quite generally using the transformation
laws of ref.~\[Gepner] that \hbox{$\alpha B_v + \beta B_s +
\gamma B_{\overline{s}}$} is a modular invariant combination of the
characters of $SO(24)$ whenever
$\alpha + \beta + \gamma = 0$. Our construction picks out precisely
the combination $2B_v - B_s - B_{\overline{s}}$. The
same combination of characters of $SO(8)$ times the singlet
of $E_8$ yields a modular invariant partition function associated
with the gauge group $E_8\times SO(8)$.

We will now describe the low-lying states of the $SO(24)$ heterotic theory
just constructed.
The only physical particle is a  massless ``tachyon''
transforming under the fundamental representation of $SO(24)$.
The vertex operator for this state is
\hbox{$e^{-\phi (z)} \psi^I(\overline{z})e^{-Q\cdot X/2}$}, where $\phi$ is
the bosonized superghost and $i\psi^I \psi^J$  yeilds the fermionic
realization of
the current algebra of $SO(24)$, with $I,J = 1,\ldots,24$.
The notation $Q\cdot X$ is as in the open string section.
Only the $r$ component of $Q$ is nonzero.

In addition, although they do not exist
as particles, it is possible to associate vertex operators with the graviton,
dilaton, antisymmetric tensor and gauge field,
whose backgrounds can appear in the
sigma model. The vertex for the gravity sector is
$e^{-\phi (z)} \psi^{\mu} (z) \overline{\partial}X^{\nu}e^{-Q\cdot X/2}$. An
$SO(24)$ gauge field is associated with the operator
$e^{-\phi (z)} \psi^{\mu} (z)(i\psi^I \psi^J)(\overline{z})e^{-Q\cdot X/2}$.

In the $(R,R)$ sector
we have fermionic backgrounds associated with the vertex operators
$e^{-\phi (z)/2} S^{\alpha} (z)\;S^A(\overline{z})e^{-Q\cdot X/2}$ and
$e^{-\phi (z)/2} S^{\dot{\alpha}} (z)\;S^{\dot{A}}(\overline{z})e^{-Q\cdot
X/2}$
where $S^{\alpha}$ ($S^{\dot{\alpha}}$) is the positive (negative) chirality
spin operator of $SO(2)$ and $S^{A}$ ($S^{\dot{A}}$) is the positive
(negative) chirality spin operator of $SO(24)$.

\chapter{Black hole geometry}

The usual charged black hole metric is
$$ds^2 = -(1- {2m\over r} + {q^2\over r^2})dt^2 + (1-{2m\over r} +
{q^2\over r^2})^{-1} dr^2.  \(stand)$$
The asymptotic flat region is at $r=\infty$, and the curvature singularity at
$r=0$.
The Reissner-Nordstrom (R-N) geometry corresponds
to the case where the two roots
$r_\pm = m \pm \sqrt{m^2-q^2}$ are both positive. The cases with $m<0$
(both roots are negative),
or the case ${m^2-q^2}< 0$ correspond to a naked singularity, since the
singularity at $r-0$ is not protected by an horizon.
The case $r_-<0<r_+<\infty$ is not usually considered since it
would correspond to imaginary $q$. This would however be a
Schwarzshild type of black hole, since one of the horizons is behind the
singularity and does not matter anymore.

We first discuss the space-time interpretation of the metric obtained in
the heterotic case,
$$ds^2 = -(1-2me^{-Qr} + q^2e^{-2Qr})dt^2 + (1-2me^{-Qr}
+ q^2e^{-2Qr})^{-1}dr^2.  \(hete) $$

The asymptotic flat region is $r=+\infty$ (since $Q$ is positive), and
the curvature singularity is at $r=-\infty$ (since $R=-g''$).
If we define $y=e^{-Qr}$,
the curvature singularity is at $y=\infty$, the asymptotic flat region is
at $y=0$, and there are two apparent singularities at
$y_{1,2}=q^{-2}(m\pm\sqrt{m^2-q^2})$. We study the case where the two roots are
distinct $y_2>y_1>0$ ( We are assuming $m>0$, and $q$ real.
 In the case $m<0$, one gets two negative
 roots, and  a naked singularity as in \(stand).
 Finally, the case when one root is positive and the other is negative
is of the Schwarzshild type, since one of the apparent singularities
is behind the asymptotic flat region.)
In the $y$ variable the metric reads
$$ds^2 = -q^2(y-y_1)(y-y_2)dt^2 +
{1\over q^2 Q^2} {dy^2 \over y^2(y-y_1)(y-y_2)}. \(rnthree)  $$
  Along the lines of ref.~\[Graves+Brill], this
 metric  can be put in conformal form
$$ ds^2 = - q^2(y - y_1)(y-y_2)(dt^2 - dr*^2)\(rnfour)$$
by the transformation
$$ r* = {1\over q^2 Q} \log y^{a_0}(y-y_1)^{a_1}
(y-y_2)^{a_2} . \(rnprob)$$
The $a_i$'s are fixed by the partial fraction decomposition
$${1\over y(y-y_1)(y-y_2)} = {a_0\over y} + {a_1\over y-y_1} +
{a_2\over y-y_2} \(decomp)$$
with $a_0^{-1}={ y_1y_2}$, $a_1^{-1}={ y_1(y_1-y_2)}$,
$a_2^{-1}={y_2(y_2-y_1)}$, and $a_0+a_1+a_2=0$.
 Notice that$a_0>0, a_1<0, a_2>0$.
The metric \(rnthree) can be put in the null form of the Kruskal-Szekeres
 variables  by introducing
$$ u_{1,2} = e^{\gamma_{1,2}(r*+t)},\qquad
v_{1,2} = e^{\gamma_{1,2}(r*-t)} \(rnsix)$$
where $\gamma_{1,2} =Qq^2(2a_{1,2})^{-1}$. This choice of $\gamma_{1,2}$
  eliminates the apparent singularity at $y=y_{1,2}$.
The null coordinates satisfy the relations
$$\eqalign{
u_1v_1 &= y^{a_0/ a_1}(y-y_1)(y-y_2)^{a_2/ a_1} \cr
u_2v_2 &= y^{a_0/a_2}(y-y_1)^{a_1/a_2}(y-y_2). \cr} \(rn12)
$$

The Penrose diagram can be built by introducing the coordinates
$\psi$ and $\xi$ defined by $u=\tan{1\over 2}(\psi+\xi)$ and
$v=\tan{1\over 2}(\psi-\xi)$.
Patching together these two patches give rise to the usual R-N
geometry, as shown in \fig{R-N}.
\insertfigure{R-N}{\vskip 4.5 truein}{Penrose diagram
for the Reissner-Nordstrom geometry.}

  The cosmological constant contribution from open string loops
gives rise to a very  similar metric
$$ds^2 = -(1- 2me^{-Qr} +2be^{-Qr/2})dt^2 + (1 - 2me^{-Qr} +
2be^{-Qr/2})^{-1}dr^2. \(rnuno) $$
The $U(1)$ string has $b=\khat/c$, while the $q=0$ limit of the
nonabelian open
string has $b = {\khat\over c}(N+2\eta)$.
In terms of the variable $y=e^{-Qr/2}$ the two apparent singularities
are at {$y_\pm= {{1\over m}(b \pm \sqrt{b^2 + m}) }$}.
For $m>0$, one always gets a positive and a negative root, therefore a
Schwarzshild-type solution.
For $m<0$ and $b<0$, the roots are both negative, and one gets a naked
singularity.
If $m<0$ and $b>0$, both roots are positive and one gets a R-N
black hole in this case.
Notice that we can get a R-N black hole even though
in this case the black hole is not charged.
 (However the case $b<0$ appears to be
 more relevant to the string solution, since $\khat > 0$ and $c<0$.)

\def\qhat{{\widehat q}}
For the general charged open string black hole, certain ranges of
parameters can lead to more interesting solutions. In this case,
using the order $q^2$ solution \(linsoln), the metric is
$$\eqalign{
ds^2 = -(1-2me^{-Qr} + &2be^{-Qr/2}+{\widehat q}^{\;2}e^{-3 Qr/2})dt^2\cr
 &+ (1- 2me^{-Qr}+ 2be^{-Qr/2}+{\widehat q}^{\;2} e^{-{3Qr/2}})^{-1}dr^2,\cr}
\(generalcubic)$$
with $\qhat = {2\khat}^{\;1/2}q/Q$.
In terms of the variable $ y= e^{-{Qr/ 2}}$ the coefficient of the
 metric is cubic, and therefore the metric has three apparent singularities at
the three roots $y_1$, $y_2$ and $y_3$, which satisfy
$$\eqalign{
y_k &= {2m\over 3{\widehat q}^{\;2}}\left\{ 1 - 2\sqrt{1-{3b{\widehat q}^{\;2}
\over 2m^2}}\ \cos\theta_k\right\}\cr
\cos(3\theta_k) = &-\left( 1-{3b{\widehat q}^{\;2}\over 2m^2}\right)^{-3/2}
 \left(1 -{9b{\widehat q}^{\;2}\over 4m^2} - {27{\widehat q}^{\;4}\over 16m^3}
\right).\cr}\(general-yks)
$$

The geometry is simplest when $b=0$, as for the $SO(2)$ case.
When $0 \le {\widehat q}^{\;4}/m^3 \le {8\over 27}$, there are three real
roots.
Since the $\theta_k$ are each separated by an angle
$2\pi/3$ when $b=0$, it follows that
for $m>0$, two roots are positive, and one negative, giving a R-N geometry.
If $m<0$, then one root is positive, giving a Schwarzshild geometry.
Outside this range of parameters, there is one real root, whose sign is
always negative, giving
a naked singularity.
If $b\ne 0$, then a range of parameters
( $m>0$, $b>0$, and ${\widehat q}^{\;2} < 0$.) exists such that
all three roots are real and positive.
The geometry of the case of three positive roots is sufficiently interesting
to merit closer attention. In fact, higher loop corrections will
lead to higher-order polynomials in $y$, as discussed in sect.~2. Therefore
multiple positive roots may be expected to occur in the generic
loop-corrected string theory.

When the three real roots are all distinct and positive,
$0<y_1<y_2<y_3<\infty$, the equivalent of
\(rnprob) is
$$ r* = a_0\log y + a_1 \log (y-y_1) + a_2\log (y-y_2) + a_3\log (y-y_3)
\(newprob)$$
where the $a_i$ coefficients are defined by the partial fraction decomposition
$$ {1\over y(y-y_1)(y-y_2)(y-y_3)} = {a_0\over y} + {a_1\over y-y_1} +
{a_2\over y-y_2} +{a_3\over y-y_3} \(dec)$$
with $a_0= -({y_1y_2y_3})^{-1}$, {\it\ etc.} and $a_0+a_1+a_2+a_3=0$.

The transformation \(newprob) allows to write the metric in the conformal
form, and the equivalent of \(rnsix) puts it in a null form. This time we
need to choose three $\gamma_i$'s, each to eliminate an apparent singularity.
$$u_i = e^{\gamma_i(r*+t)}, \qquad v_i = e^{\gamma_i(r*-t)} \(rnseven)$$
where $\gamma_i=(2a_i)^{-1}, i=1,2,3$. To understand the geometry of
each patch,  the relations
$$\eqalign{
u_1v_1 &= y^{a_0/ a_1}(y-y_1)(y-y_2)^{a_2/ a_1}(y-y_3)^{a_3
/ a_1} \cr
u_2v_2 &= y^{a_0/ a_2}(y-y_1)^{a_1/ a_2}(y-y_2)(y-y_3)^{a_3
/ a_2} \cr
u_3v_3 &= y^{a_1/ a_3}(y-y_1)^{a_1/ a_3}(y-y_2)^{a_2/ a_3}
(y-y_3) \cr} \(patches) $$
are helpful.
Since the $a_i$'s have alternate signs, the ratios ${a_2/ a_3}$ and
${a_2/a_3}$ are negative. Therefore $u_1v_1=0$ at $y=y_1$ and
$u_1v_1=\infty$ at $y=y_2$, and similarly in the other patches.
Every time we cross a line $y=y_i$, the space and time coordinate
flip role. So in this case the singularity at $y=\infty$
is space-like again, as in the Schwarzschild case.
 The final Penrose diagram is the two dimensional lattice
illustrated in \fig{Penrose}.
When higher order loop corrections are included, one can obtain even more
positive roots in principle, leading to more intricate geometries.
\insertfigure{Penrose}{\vskip 6 truein}{Penrose diagram for the charged
open string. The dotted areas are not part of space-time.}

\chapter{General charge and mass formulas}

The computation of the charges is a straightforward application
of Gauss's law.

First, consider the heterotic string effective action \(Shet).
Adding a source $-\int d^2 x \sqrt{-G} J_\mu A^\mu$ to the action \(Shet),
shows that the conserved current density is
$$
J_\mu = \nabla^\nu\left(e^{-2\Phi}F_{\nu\mu}\right). \(hetcurrent)
$$
Charge is the integral of $J_t$ over $r$, which is just the value at
$r\rightarrow\infty$ of $e^{-2\Phi}F_{tr}$. This quantity is independent
of $r$ by the equations of motion. Then the charge of the black hole is
$$
{\cal Q} = \int dr \; J^t(r) =
\left(e^{-2\Phi}f(r)\right)_{r\rightarrow\infty}
= \sqrt{2}Q q e^{-2\Phi_0}  \(syQhet)
$$
using the solution of \(mm2) together with $Q = \sqrt{8\over \alpha'}$.

The Born-Infeld case proceeds similarly. In this case, $A^\mu$ was rescaled
by a power of $\alpha'$, giving it dimensions of length, so the appropriate
current coupling is \hbox{$-(2\pi\alpha')^{-1/2}\int d^2 x
\sqrt{-G} \Tr(J_\mu A^\mu)$}. Then the current derived from \(nonabelian) is
$$
J_\mu = \sqrt{2\pi\alpha'}\kappa\nabla^\nu\left[e^{-\Phi}\sqrt{-\det(G+F)}
\left({F\over G-F^2}\right)_{\nu\mu}\right]
\(BIcurrent)
$$
where the {\it r.h.s.} of \(BIcurrent) is connected to (3.2) via Bianchi
identities as in \[ACNY].
The quantity in brackets is  constant for a solution to the equations
of motion, and gives the physical
charge
$$
{\cal Q} = \sqrt{2\pi\alpha'}{\kappa} q e^{-\Phi_0} =
8\lambda q \sqrt{2\pi\over\alpha'} e^{-\Phi_0}\(BIcharge)
$$
of the black hole, where $\lambda$ is the dimensionless parameter
defining the normalization of the one-loop partition function in (A.1)
of the appendix.

We now discuss the masses.
In general relativity the mass is determined from the asymptotic behaviour of
the gravitational field. Any gauge field or
loop induced  cosmological constant
that might be present  could change the mass via its
contribution to the behaviour of the gravitational field at infinity.
 In ref.~\[Witten],  the mass of the black hole was computed using the
standard ADM prescription, namely
by perturbing  the black hole solution around asymptotically flat space with
a linear dilaton. We could  follow  the ADM procedure as well. We would then
find  that in all the cases discussed in this paper the mass stays unchanged,
 and it is the same as in \[Witten]. The reason is that the changes
in the  dilaton stress-energy tensor
$$T^\Phi_{\mu\nu} = e^{-2\Phi}(\beta^G_{\mu\nu} +
 {1\over 2}g_{\mu\nu}\beta^\Phi) \(stress)$$
 coming
from loop corrections, as in (2.10), are cancelled against the change in the
 metric, as in (2.11).
 However we will not follow the ADM derivation here, but  choose
to start from a quasi-local mass formula, an  approach that  might be of
 interest in its own merits. We will find of course the same results,
and even the details of the calculation are practically the same.
 In any number of dimensions, it is actually possible
to give a local definition of the energy density, provided one considers only
spherically symmetric solutions\[MTW]. The physical reason is that there are
no gravitational S-waves, so one can unambiguously define the mass enclosed
in a sphere, and construct a conserved matter stress-energy tensor. In $D=2$,
there are never any gravity waves. We can think of this as a degenerate
spherical case, and look for a mass density function whose integral
from $-\infty$ to $r$ gives the mass to the left of $r$. We will find a
conserved vector, whose time component is the mass density, and show that
its integral reproduces the ADM results. The main point is that in two
 dimensions one does not need to perturb about flat space-time
in order to define a cunserved current.discussion. Indeed,
 taking inspiration from ref.~\[FMP], one can define
$$
S_{\mu} = {{\sqrt{\alpha'}}\over2}\epsilon^{\lambda\nu}T^{m}_{\lambda\mu}
\partial_\nu\Phi  \(polc1)$$
Since $R_{\mu\nu} - \coeff12 RG_{\mu\nu}$
vanishes identically in two dimensions,
the equations of motion imply $T^{\Phi}_{\mu\nu} + T^m_{\mu\nu} = 0$,
 (we omit the tachyon or include it in $T^m_{\mu\nu}$).
Therefore \(polc1) can be also written as
$$
S_\mu = -{{\sqrt{\alpha'}}\over2}\epsilon^{\lambda\nu}T^{\Phi}_{\lambda\mu}
                                                \partial_\nu \Phi. \(polc)
$$
Starting from the definition \(polc) we show that
in all the cases we examine in this paper $S^\mu$ is a total derivative,
 \hbox{$S^\mu = \epsilon^{\mu\nu}\partial_\nu\omega$} with
an appropriate $\omega$, so that $S_\mu$ is a conserved current.

 We will first compute $\omega$ for the
standard blackhole \[Witten].
 We find convenient to work in the target space
conformal gauge $G_{\mu\nu} = g \eta_{\mu\nu}$
 so that our variables are $t$ and $r*$, as in the previous section. Then
the components of $T^\phi_{\mu\nu}$ (with no loop
correction included so far) are
$$\eqalign{
T^{\Phi}_{tt} &= e^{-2\Phi}\left(2 {\dot{g} \over g}\dot{\Phi}
                      +2{g' \over g}\Phi' - 4{\dot{\Phi}}^2
                      +4{\Phi'}^2 - 4\Phi'' + cg\right),\cr
T^{\Phi}_{r*r*} &= e^{-2\Phi}\left(2 {\dot{g} \over g}\dot{\Phi}
                      +2{g' \over g}\Phi' + 4{\dot{\Phi}}^2
                      -4{\Phi'}^2 - 4\ddot\Phi - cg\right),\cr
T^{\Phi}_{tr*} &= e^{-2\Phi}\left(2{\dot{g}\over g} \Phi'+ 2{g'\over g}
                      \dot{\Phi} - 4{\dot{\Phi}}'\right),\cr} \(Tgdef)
$$
where the  primes denote  derivatives with respect to $r{*}$, {\it i.e.}
$\Phi' = g{d\over dr}\Phi$. In this case
$$
\omega = {\sqrt{\alpha'}\over {g}}e^{-2\Phi}\left({\dot{\Phi}}^2 -
    {\Phi'}^2\right) + {4\over {\sqrt{\alpha'}}}e^{-2\Phi}. \(Mdef)
$$
This implies that $\partial_{\mu}S^{\mu} = 0 $,
so that $S^\mu$ is a conserved current. The component $S^t$ represents the
conserved mass density, and $S^{r*}$ measures its flow in non-static
situations.
In an asymptotically flat case with a linear dilaton background,
$S^\mu$ reduces to the linearized $S^\mu$
defined in  ref.~\[Witten]. Therefore it will reproduce the ADM mass result
in that case, as we will check.

The idea of using the scalar field $\omega$ to define a local mass
has been discussed extensively by Poisson and Israel\[PI].
For a static solution, $S^{t}$ defines
a conserved mass \hbox{$M(r*) = \int^{r*} d{r*}\ S^{t}(r*)$.}
Therefore, up to an integration constant, \hbox{$M(r*) = \omega(r*)$} defines
the mass contained inside a distance $r*$. For a non-singular
solution, the integration constant would be unambiguous\[MTW],
and would correspond to taking the lower limit to $-\infty$ in this integral.
However, we are interested in a solution which is singular at that point,
so we instead define the black hole mass to be \hbox{$M = \omega(\infty) -
\omega_0(\infty)$}, where $\omega_0$ is the mass function for ``empty space,''
\ie\ for parameters $m=q=0$ in the black hole metric. In the absence of a
cosmological constant, $\omega_0 = 0$.

For the static solutions we have considered, $M$ can be expressed as an
integral
over the matter stress tensor component $T^m_{tt}$, giving further evidence
that this definition of the mass is physically what one would expect.
Note that evaluating \(polc1) for $\mu=t$ gives
$$
\omega'
= {\sqrt{\alpha'}\over 2g}(-T^m_{tt}\Phi' + T^m_{tr*}\dot{\Phi}). \(newt1)
$$
Using the dilaton ansatz
\hbox{$\Phi = \Phi_0 - {2\over {\sqrt{\alpha'}}}r$}, we obtain
$$
M = \int d{r*}\ T^m_{tt}(r*) = \int dr\ g^{-1} T^m_{tt}.    \(newt2)
$$
This shows that the mass is determined by the energy
density of the matter source. (The expression \(newt2) is an alternative
way to verify the normalization of $S^\mu$.)

Evaluating the mass formula \(Mdef) on the classical solution one can relate
the parameter $m$ of the black hole solution to the physical mass. The
result is
$$
M = {8\over {\sqrt{\alpha'}}} m e^{-2\Phi_0}     \(BigM)
$$
This expression is in agreement with ref.~\[Witten]'s definition of the mass if
we take $2m = 1$. This agreement with the ADM result is a sign that our
definition of the mass is physically reasonable.

  Given  the black hole mass one can now determine
 the black hole entropy, which is given by
$$
S = \beta_H M + \ln Z = (\pi \sqrt{\alpha'})
({8\over {\sqrt{\alpha'}}} me^{-2\Phi_0})
  = 8\pi m e^{-2\Phi_0}= \pi \sqrt{\alpha'} M.    \(BHS)
$$
Here $\beta_H $ is the inverse Hawking temperature of the two dimensional
black hole and $\ln Z$ is the Euclidean action which can be shown to vanish for
the ordinary closed string case.
(This expression has also been obtained by \[Trevidi].)

As derived, the mass formula \(Mdef) valid only for the
ordinary closed-string action \(closedaction).
However, the same principles will apply in general.
 Other theories  will have new contributions to the stress-energy tensor,
 which must be included in eq.~\(polc), leading to a modified
 mass function $\omega$. For example, the loop corrected effective
action to the two dimensional closed string, including
the cosmological constant term \(cosmo) for $n=1$,
yields a mass formula
$$\eqalign{
\omega &= {e^{-2\Phi}\over g}({\dot{\Phi}}^2 - {\Phi'}^2)\sqrt{\alpha'}
    + {4 \over {\sqrt{\alpha'}}}e^{-2\Phi}
    - {1\over 2} {\sqrt{\alpha'}} \Lambda \Phi \cr
  &= {8\over\sqrt{\alpha'}}me^{-2\Phi_0} + {\sqrt{\alpha'}\over2}\Lambda\Phi_0.
    \cr}  \(closedloopmass)
$$
Subtracting $\omega_0 = {1\over2}\sqrt{\alpha'}\Lambda\Phi_0$ gives the
same value \(BigM) as the standard black hole.
In the open string case, the loop-induced cosmological constant modifies
the mass formula as follows:
$$
\omega = {e^{-2\Phi}\over g}({\dot{\Phi}}^2 - {\Phi'}^2)\sqrt{\alpha'}
    + {4 \over {\sqrt{\alpha'}}}e^{-2\Phi}
    + {1\over 2} {\sqrt{\alpha'}} \Lambda e^{-\Phi}.  \(openmass)$$
However, the actual value of $M$ is independent of
$\Lambda$, and is identical with the standard result \(BigM). (In this case,
as in all of the following open string results, $\omega_0 = 0$, and no
subtraction is needed to obtain the mass.)

Finally, we turn to the mass formulas of the charged black holes.
In the heterotic case we obtain
$$
\omega = {4\over Qg}e^{-2\Phi}\left({\dot{\Phi}}^2 -
 {\Phi'}^2\right)
    + Qe^{-2\Phi}
    + {1\over 2Q}{\cal Q}^2 e^{2\Phi} \(MHet)
$$
while the open string Born-Infeld action yields
$$\eqalign{
\omega =&\ {\sqrt{\alpha'}\over g}e^{-2\Phi}\left({\dot{\Phi}}^2 -
 {\Phi'}^2\right)
    + {4 \over {\sqrt{\alpha'}}}e^{-2\Phi}
    + \kappa\sqrt{\alpha'} \left(1-\eta-{N\over2}\right) e^{\Phi}  \cr
&- \kappa\sqrt{\alpha'}e^{\Phi}
\left(1 + {{\cal Q}^2 e^{2\Phi}\over 2\pi\alpha'\kappa^2}\right)^{1/2}
+ {{\cal Q}\over {\sqrt{2\pi}}}
\sinh^{-1}\left(e^{\Phi}{{\cal Q}\over \sqrt{2\pi\alpha'}\kappa}\right).
 \cr}\(BIM)
$$
In either case, evaluating these formulas gives the mass $2Qme^{-\Phi_0}$
 but for the
heterotic case $Q = \sqrt{{8 \over \alpha'}}$ whereas in the bosonic
case $Q= {4\over {\sqrt{\alpha'}}}$.

\chapter{Conclusions}
In this paper we have investigated more general black hole solutions
to two dimensional string theory. The modifications have been obtained
introducing  gauge fields  by coupling them to the
boundary of  open strings. We have shown how these corrections can be
derived in the context of two dimensional strings by applying
the techniques of boundary operators \[CLNY2] and BRST invariance.
Our results hold only in the leading order in $\alpha'$.
 Since the completion of this work, some
further analysis of the $2d$ open string has appeared\[BerKut], which
possibly leads to an exact treatment.

Another way to couple a gauge field to strings is to construct a heterotic
string. We found such a theory with an $E_8\times SO(8)$ or
$SO(24)$ gauge group. Applying background field methods
allowed us to find a second form of charged black hole solutions. These are
pure closed string tree level solutions, although the space-time metric
has the same form as the loop-corrected ones.
 Our solution appears to be different
from those recently discussed in  \[ILS, Kumar].

The solutions have geometries which can be identified with Schwarzshild or
Reissner-Nordstrom black hole solutions. Both the open and heterotic string
theory can give rise to either class of solutions, depending on the range of
charge and mass parameters. Models with a cosmological constant can lead to
a Reissner-Nordstrom geometry even in the absence of charge.

For a certain range of parameters of the charged open string,
there can be three event horizons.
These solutions have a very interesting geometry, which can be described as a
two dimensional lattice of black holes. In this case, the Penrose diagram tiles
the plane. Multiple event horizons should be the generic case when higher order
loop corrections are included, leading to very intricate geometries in general.

\ack
We wish to thank D. Kutasov, S. Trivedi, J.Polchinski
and especially  Z.Qiu  for useful discussions.
 S.~Yost thanks the
Institute for Advanced Study for its hospitality during the completion of
this work.

\append{Open String Partition Function and Factorization}

In this appendix, we check that the boundary state plays the usual
role\[CLNY2] in factorizing the open string partition function.
This fixes its normalization, once the one-loop open string vacuum
amplitude is given.

The one-loop open string vacuum amplitude may be expressed as
a product of partition functions for the string modes,
$$
Z = \lambda^2\int_0^{\infty} {d\tau\over\tau}
Z_X(\tau)Z_{\rm ghost}(\tau) \(Zfactors)
$$
where we are considering a strip of width $\pi$, whose ends are identified
after a propagation time $2\pi i \tau$. The open-string one loop coupling
constant is $\lambda^2$, the square of the tree-level coupling.
Choosing a convenient normalization,
$$
Z_X(\tau) = {8\pi\alpha'}\int {d^2 P\over (2\pi)^2} \Tr e^{2\pi i\tau H} =
     (-\pi i\tau)^{-1}\eta^{-2}(\tau)         \(Zrt).
$$
where the open string Hamiltonian is\[GSW]
$$
H = \alpha' P^2 + \sum_{n=1}^\infty  \alpha_{-n}\cdot
{\widetilde \alpha}_{-n} - {1\over 12}       \(Hopen)
$$
including the normal-ordering constant $-1/24$ per boson\[GSW],
and
$$\eta(\tau) = (-i\tau)^{-1/2}\eta(-1/\tau) =
        e^{\pi i\tau/12}\prod_{n=1}^{\infty} (1-e^{2\pi in\tau})$$
is the Dedekind eta function. The momentum
$P_r$ includes the shift by $-iQ/2$ explained in sect.~3, but this does not
affect $Z_X$ since $P$ is integrated from $-\infty$ to $\infty$.
The ghost partition function is simply\[GSW]
$$
        Z_{\rm ghost}(\tau) = \eta^2(\tau).     \(Zghst)
$$

A Jacobi transformation $\tau\rightarrow -1/\tau$ transforms the partition
function to the closed string channel, where it may be interpreted as an
overlap of boundary states. The result of the Jacobi transformation is
$$\eqalign{
Z_X(\tau) &= {1\over\pi} a^{-1/6}\prod_{n=1}^{\infty} (1-a^{2n})^{-2}\cr
Z_{\rm ghost}(\tau) &= -{\ln a\over\pi} a^{1/6}\prod_{n=1}^{\infty}
        (1-a^{2n})^2\cr} \(closedchannel)
$$
where $a = e^{-\pi i/\tau}$ is the modulus of an annulus conformally equivalent
to the tube of length $-\ln a$. The total partition function is then
$$
Z = \lambda^2 \int_0^1{da\over a}.
$$
This has a logarithmic divergence at $a\rightarrow 1$ from the ``tachyon''
tadpole. In the presence of background fields, $Z$ is multiplied by\[CLNY1]
$$
        {\det[G(x)+F(x)]}.      \(Ffactor)
$$

Each of the partition function factors in \(Zfactors) may be individually
expressed as an overlap of the appropriate boundary states in the closed
string channel. The ghost partition function is factorized in ref.~\[CLNY2],
and it is easy to check that
$$
Z_X(\tau) = \bra{B} a^{L_0 + {\widetilde L}_0 }\ket{B}_X .
\(Zfactorized)
$$
Therefore, the boundary states play the same role in factorizing the
loop amplitude as they do in the critical case. The normalization of
the boundary state is fixed by the normalization of the one loop partition
function via \(Zfactorized), which in turn fixes the normalization of the
loop corrections to the beta functions.
%
%
%
\refdef{Witten}{E. Witten, ``On String Theory and Black Holes\rlap,''
 IAS preprint IASSNS--HEP--91/12 (1991)}
\refdef{blackholes}{R. Dijkgraaf, H. Verlinde and E. Verlinde, ``String
 Propagation in a Black Hole Geometry\rlap,'' Princeton preprint PUPT--1252
 (1991)}
\refdef{MTW}{C. Misner, K. Thorne and J.A. Wheeler, {\it Gravitation}
 (W.H.~Freeman \&\ Co., 1973)}
\refdef{CLNY1}{C. Callan, C. Lovelace, C. Nappi and S. Yost, Nucl.~Phys.~B288
 (1987) 525}
\refdef{CLNY2}{C. Callan, C. Lovelace, C. Nappi and S. Yost, Nucl.~Phys.~B293
 (1987) 83}
\refdef{CLNY3}{C. Callan, C. Lovelace, C. Nappi and S. Yost, Phys.~Lett.~206B
 (1988) 41, Nucl.~Phys.~B308 (1988) 221}
\refdef{BI}{M. Born and L. Infeld, Proc.~Royal Soc.~144 (1934) 425\hfill\break
 E. Fradkin and A. Tseytlin, Phys.~Lett.~163B (1985) 123}
\refdef{ACNY}{A. Abouelsaood, C. Callan, C. Nappi and S. Yost,
 Nucl.~Phys.~B280 [FS18] (1987) 599}
\refdef{yost}{S. Yost, Nucl.~Phys.~B321 (1989) 629}
\refdef{DK}{J. Distler and H. Kawai, Nucl.~Phys.~B321 (1989) 509}
\refdef{Mandal}{G. Mandal, A. Sengupta and S. Wadia, ``Classical Solutions of
 Two Dimensional String Theory\rlap,'' IAS preprint
 IASSNS-HEP-91/10 (1991)}
\refdef{Dealw}{S.P. De Alwis and J. Lykken, ``2d Gravity and the Black Hole
Solution in 2d Critical String Theory'', Fermilab-pub-91/198-T.}
\refdef{Kumar}{S. Pratik Khastgir and Alok Kumar, Bhubaneswar preprint, 1991}
\refdef{Verl}{R. Dijkgraaf, H. Verlinde and E. Verlinde, ``String Propagation
in a Black Hole geometry'', IASSNS-HEP-91/22}
\refdef{CFMP}{C. Callan, D. Friedan, E. Martinec and M. Perry, Nucl.~Phys.~B262
 (1985) 593}
\refdef{CG}{C. Callan and Z. Gan, Nucl.~Phys.~B272 (1986) 647}
\refdef{spec}{S. Das and A. Jevicky, Mod.~Phys.~Lett.~A5 (1990)
1639\hfill\break
 J.~Polchinsky, Nucl.~Phys.~B346 (1990) 253}
\refdef{Love}{C. Lovelace, Nucl.~Phys.~B273 (1986) 413}
\refdef{FS}{W. Fischler and L. Susskind, Phys.~Lett.~173B (1986) 262}
\refdef{FMS}{D. Friedan, E. Martinec and S. Shenker, Nucl.~Phys.~B271 (1986)
 93}
\refdef{tachyonbeta}{S. Das and B. Sathiapalan, Phys.~Rev.~Lett.~56 (1986)
 2664\hfill\break C. Itoi and Y. Watabiki, Phys.~Lett.~198B (1987) 486}
\refdef{GSW}{M. Green, J. Schwarz and E. Witten, {\it Superstring Theory},
 vol.~1 (Cambridge University Press, 1987)}
\refdef{SZ}{W. Siegel and B. Zweibach, Nucl.~Phys.~B263 (1986) 105
\hfill\break T. Banks and M.Peskin, Nucl.~Phys.~B264 (1986) 513}
\refdef{opengroup}{N. Marcus and A. Sagnotti, Phys.~Lett.~119B (1982) 97\hfill
 \break M. Green, J. Schwarz and E. Witten, {\it Superstring Theory}, vol.~2
(Cambridge University Press, 1987) chap.~8.1.2}
\refdef{dAPS}{S. deAlwis, J. Polchinski and R. Schimmrigk, Phys.~Lett.~B218
 (1989) 449}
\refdef{ILS}{N. Ishibashi, M. Li and A. Steif, UCSB preprint
UCSBTH--91--28--REV
 (1991)}
\refdef{FMP}{W. Fischler, D. Morgan and J. Polchinski, Phys.~Rev.~D41 (1990)
  2638, Phys.~Rev.~D42 (1990) 4042}
\refdef{Trevidi}{S. Trevidi, private communication}
\refdef{Gepner}{D. Gepner, Nucl.~Phys.~B296 (1988) 757}
\refdef{P+C}{J. Polchinsky and Y. Cai, Nucl.~Phys.~B296 (1988) 91}
\refdef{Graves+Brill}{J.C. Graves and D.R. Brill, Phys.~Rev.~120 (1960) 1507}
\refdef{Carter:Kerr}{B. Carter, Phys.~Rev.~141 (1966) 1242}
\refdef{Carter:e2=m2}{B. Carter, Phys.~Lett.~21 (1966) 423}
\refdef{Hawking+Ellis}{S.W. Hawking and G.F.P. Ellis, {\it The Large Scale
 Structure of Spacetime}\ (Cambridge Univ.~Press, 1973)}
\refdef{ThornSFT}{C. Thorn, Phys.~Rep.~175 (1989) 1}
\refdef{ChodosThorn}{D. Fairlie, unpublished \hfill\break
 A. Chodos and C. Thorn, Nucl.~Phys.~B72 (1974) 509}
\refdef{CHSW}{P. Candelas, G. Horowitz, A. Strominger and E. Witten,
 Nucl.~Phys.~B258 (1985) 46}
\refdef{ChanPaton}{J. Paton and H. Chan, Nucl.~Phys.~B10 (1969) 516}
\refdef{KS}{D. Kutasov and N. Seiberg, Princeton preprint PUPT-1193, RU-90-41}
\refdef{KRT}{Kounnas, B. Rostand and E. Tomboulis, UCLA preprint UCLA/91/TEP 1}
\refdef{PI}{E. Poisson and W. Israel, Phys.~ Rev.~D41 (1991) 1796}
\refdef{Kutcom}{D. Kutasov, private communication}
\refdef{Polcom}{J. Polchinski, private communication}
\refdef{DH}{L. Dixon and J. Harvey, Nucl.~Phys.~B274 (1986) 93}
\refdef{SW}{N. Seiberg and E. Witten, Nucl.~Phys.~B276 (1986) 272}
\refdef{BerKut}{M. Bershadsky and D. Kutasov, Princeton preprint PUPT-1283}
\references{}
\figures{}
\end